\documentclass[prb,amsmath,amssymb,superscriptaddress,twocolumn]{revtex4}
\usepackage{amsmath}
\usepackage{braket}
\usepackage{graphicx}
\usepackage{multirow}
\usepackage{bbm,color,ulem}
\DeclareMathOperator{\sinc}{sinc}
\allowdisplaybreaks

\begin{document}
\title{Error correction for gate operations in systems of exchange-coupled singlet-triplet qubits in double quantum dots}
\author{Donovan Buterakos}
\author{Robert E.\ Throckmorton}
\author{S.\ Das Sarma}
\affiliation{Condensed Matter Theory Center and Joint Quantum Institute, Department of Physics, University of Maryland, College Park, Maryland 20742-4111 USA}
\date{\today}
\begin{abstract}
We present a scheme for correcting for crosstalk- and noise-induced errors in exchange-coupled singlet-triplet semiconductor
double quantum dot qubits.  While exchange coupling allows the coupling strength to be controlled independently of the intraqubit
exchange couplings, there is also the problem of leakage, which must be addressed.  We show that, if a large magnetic field difference
is present between the two qubits, leakage is suppressed.  We then develop pulse sequences that correct for crosstalk- and noise-induced
errors and present parameters describing them for the 24 Clifford gates.  We determine the infidelity for both the uncorrected
and corrected gates as a function of the error-inducing terms and show that our corrected pulse sequences reduce the error by
several orders of magnitude.
\end{abstract}
\maketitle

\section{Introduction}

Semiconductor-based electron spin qubits are one of several platforms that are currently being actively investigated,
both theoretically and experimentally, with the goal of eventually building a quantum computer.  These types of qubits
are formed from one or more quantum dots with a single electron trapped inside each.  A number of different varieties
of semiconductor electron spin qubits exist, such as the single-spin exchange qubit\cite{LossPRA1998,HuPRA2000,HuPRA2001,NowackScience2011,PlaNature2012,BraakmanNatNano2013,PlaNature2013,VeldhorstNatNano2014,OtsukaSciRep2016,ItoSciRep2016},
the singlet-triplet double-dot qubit\cite{LevyPRL2002,PettaScience2005,FolettiNatPhys2009,VanWeperenPRL2011,MauneNature2012,ShulmanScience2012,DialPRL2013,ShulmanNatCommun2014,ReedPRL2016,MartinsPRL2016},
the triple-dot exchange-only qubit\cite{DiVincenzoNature2000,MedfordNatNano2013,MedfordPRL2013,EngSciAdv2015,ShimPRB2016},
and a double-dot ``hybrid'' qubit with three electrons total\cite{ShiPRL2012,KimNature2014,KimNPJQI2015}.  These various
spin qubit platforms have been realized in both Si-based and GaAs-based architectures.  Other quantum computing platforms
include superconducting transmons and trapped ions.  While other platforms currently have better coherence times and higher
fidelity, allowing a larger number of gates to be performed, semiconductor electron spin qubits allow for faster gates and are
compatible with the existing semiconductor industry, thus allowing easier scaling up for eventual real applications beyond just
laboratory demonstrations of principles.  The greatest challenge for semiconductor-based qubits currently is thus improving
coherence time and fidelity.  Much progress has been made on this issue, with experiments on singlet-triplet qubits
reporting fidelities as high as 99\% for single-qubit gates and 90\% for two-qubit gates.  This still falls short of the goal
of at least 99\% required to implement surface codes\cite{FowlerPRA2012}, let alone the 99.99\% fidelity in all operations
required before other error-correcting techniques can begin to be implemented.  This unfortunately compares unfavorably with
superconducting qubits, which claim fidelities of over 99.9\% for single-qubit gates and over 99\% for two-qubit gates\cite{BarendsNature2014},
and with ion trap qubits, which claim similar fidelities\cite{BenhelmNatPhys2008}.  However, both types of qubits have gate times
on the order of $\mu$s, compared to ns for semiconductor-based electron spin qubits.  The fact that so much experimental
progress has been made in semiconductor spin qubits over the last decade is therefore promising.  The subject is currently
highly active with around ten large interdisciplinary groups being involved in spin qubit experiments all over the world.

We will focus on the singlet-triplet qubit in this work.  These qubits consist of two quantum dots, each with a single electron
trapped inside, coupled via exchange coupling and subject to a magnetic field gradient (i.e., magnetic fields that differ at
each dot).  The exchange coupling can be varied purely electrically, allowing for fast manipulation of the qubit.  Typically,
the field gradient is created by either depositing a micromagnet nearby or by polarizing the nuclear spins (if possible) and cannot
be changed quickly, so it is held constant.  However, a recent work\cite{CollardArXiv} attempts to realize this gradient electrically
by tuning the effective $g$ factor within each quantum dot, which would allow for fast manipulation of this gradient.  We will
assume, however, that this gradient is held constant, as this is the case in most experiments.  We will see later
that the form of the effective Hamiltonian for this qubit restricts us to rotations about axes in one quadrant of the $xz$ plane,
and furthermore does not allow pure $x$ or $z$ rotations, meaning that rotations about other axes must be performed with complex
pulse sequences.  Not only does this mean that most rotations require more complex pulse sequences to perform, but also that we
cannot implement common error correction schemes such as the NMR-inspired Hahn echo technique or its generalization, the Carr-Purcell-Meiboom-Gill
(CPMG) technique\cite{WitzelPRL2007,WitzelPRB2007,LeePRL2008}.  Other techniques are thus required, and have in fact been developed.

We will assume, consistent with the experimental situation, that our qubits are subject to two types of noise---magnetic field noise,
present in the field gradient, and charge noise, which manifests as noise in the exchange coupling.  We will assume throughout this work
that the noise is quasistatic, which is often a good approximation.  While the idea of dynamical decoupling in semiconductor spin qubits
through specially-designed pulse sequences to improve retention of the qubit state has been around for years\cite{BluhmPRL2010,BluhmNatPhys2011,SergeevichPRA2011,ShulmanNatCommun2014,MuhonenNatNanotechnol2014,MalinowskiNatNanotechnol2017},
the problem of error correction in the performance of gate operations was first considered in Ref.\ \onlinecite{WangNatComm2012}, in
which the technique of Soft Uniaxial Positive Control for Orthogonal Drift Error (\textsc{supcode}) was introduced, and
further developed in later work\cite{WangPRA2014,ThrockmortonPRB2017}.  This technique makes use of square pulses (i.e., a piecewise
constant $J$) both to implement gates and to correct errors to arbitrary order.  An even more recent work\cite{ZengArXiv} considers
error correction to arbitrary order using smooth pulses, but considers only magnetic field noise.

These works considered singlet-triplet qubits in isolation, without any interqubit coupling.  To build a practical quantum computer, one
must couple qubits together so that multiqubit gates can be performed as well.  The ability to perform arbitrary single-qubit gates and at
least one two-qubit gate are necessary for universal quantum computation.  The effects of noise on such two-qubit operations
has in fact been considered in several works\cite{WuPRB2017,WangPRA2014,WangNPJQI2015,CalderonVargasPRL2017} (the last of these is actually
platform-independent).  Such coupling, unfortunately, also introduces a new challenge---crosstalk.  The fact that the qubits are coupled
means that, while a single-qubit operation is being performed, the changes in the qubit's state also cause unintended changes in nearby qubits.
As a result, we need to correct for crosstalk as well.

There are (at least) two ways to couple singlet-triplet qubits.  One method is through capacitive coupling, in which two qubits are coupled
via interaction of their dipole moments.  The singlet state of a singlet-triplet qubit possesses a nonzero dipole moment, but not the triplet
state, resulting in a state-dependent electrostatic interaction.  This interaction is empirically found\cite{ShulmanScience2012} to be proportional
to the exchange couplings in the two qubits.  Correcting for both noise- and crosstalk-induced error in single-qubit gate operations was the
subject of a recent work of ours\cite{ButerakosPRB2018}, which uses a technique based on \textsc{supcode} to cancel errors in single-qubit
operations to leading order.  The other coupling method, which is also used to couple singlet-triplet qubits, is exchange coupling, in which
one of the spins in one qubit is coupled to a spin in the other qubit.  This has the advantage of allowing independent control of the intraqubit
exchange couplings and the interqubit coupling.  However, this also allows for leakage of the system out of the computational subspace; this
coupling could, for example, put the system into the state, $\ket{\uparrow\uparrow\downarrow\downarrow}$.  The subject of correcting errors
in single-qubit gates performed on one of a pair of exchange-coupled singlet-triplet qubits is the focus of the present work.

The system under consideration is a pair of exchange-coupled singlet-triplet qubits with identical magnetic field gradients, but
subject to differing overall magnetic fields.  As has been shown, it is not necessary that the magnetic field gradients of the qubits
be identical, but we shall assume that they are for simplicity.  The differing overall magnetic fields are crucial to allowing us
to tune the energy cost of entering the two possible leakage states, $\ket{\uparrow\uparrow\downarrow\downarrow}$ and $\ket{\downarrow\downarrow\uparrow\uparrow}$.
We split the Hamiltonian of this system into two parts, one that only connects states within the computational subspace and the
magnetic field-induced separation of the leakage states, and those which are responsible for leakage.  We then apply perturbation
theory to show that a sufficiently large magnetic field can suppress leakage and argue that we can ignore the leakage terms for
the remainder of this work.  We then outline our approach for developing pulse sequences that correct for noise- and crosstalk-induced
errors.  This approach is similar to that of Ref.\ \onlinecite{ButerakosPRB2018},where the case of capacitive interqubit coupling
was addressed (in contrast to the current work addressing interqubit exchange coupling).  It consists of following the na\"ive
single-qubit gate with identity operations on both qubits arranged in such a way as to cancel the error in the gate to first order.
We add these pulses in ``blocks'' consisting of two pulses on each qubit, arranged such that both pairs have the same duration.
This differs from Ref.\ \onlinecite{ButerakosPRB2018} since, in this previous work, the ``blocks'' instead consisted of two pulses
on one qubit and a single pulse on the other.  We find that the arrangement of the ``blocks'' used in this work allow for shorter
pulse sequences, both in time and in number of ``blocks'' needed.  We obtain the parameters needed to correct errors in the na\"ive
pulse sequences introduced in Ref.\ \onlinecite{ThrockmortonPRB2017} and show that these sequences reduce the errors (infidelities)
in the gates by {\it several orders of magnitude}.  However, we find that, for low noise and low coupling, the error still scales linearly,
only transitioning to a quadratic dependence for larger values.  This is due to some residual leakage, since we neglect the effect
of the terms that cause it.  We then point out, however, that the range over which we obtain this linear dependence can be reduced
by increasing the magnetic field difference between the qubits.

The rest of the manuscript is organized as follows.  Section II introduces the Hamiltonian that we use in the rest of this
work and quantifies the magnitude of the effect of leakage out of the computational subspace.  In Sec.\ III, we present the
formulas for the first-order error due to crosstalk and noise.  Section IV describes the methods that we use to correct for
these errors, and we describe our results in Sec.\ V.  We give our conclusions in Sec.\ VI, and present the detailed numerical
parameters for our error-corrected pulse sequences in the Appendix.

\section{Derivation of the Hamiltonian}

In the singlet-triplet encoding scheme, each qubit is encoded in the spin degrees of freedom of two electrons bound in a
pair of quantum dots. An external magnetic field is applied along the $z$-axis, conserving the $z$-component of the
total spin $S_{1z}+S_{2z}$, and thus the computational space is given by the space with $S_z=0$. We define the basis states
as $\ket{0}=\frac{\ket{\uparrow\downarrow}+\ket{\downarrow\uparrow}}{\sqrt{2}}$, $\ket{1}=\frac{\ket{\uparrow\downarrow}-\ket{\downarrow\uparrow}}{\sqrt{2}}$,
though other conventions exist. The effective Hamiltonian for a single qubit in this basis is given by
\begin{equation}
H=hX+J(t)Z,\label{eqn:singleh}
\end{equation}
where $X$ and $Z$ are the Pauli matrices, $h$ is the magnetic field gradient between the two quantum dots, and $J$ is the
exchange coupling.  The strength of the magnetic field gradient can only be varied slowly compared to the qubit coherence time,
and so we treat $h$ as constant, controlling the qubit only by varying the strength of $J$. Critically, the sign of $J$
remains the same, meaning that only forward rotations about axes lying in one quadrant of the $xz$-plane can be performed.

We now consider a system of two singlet-triplet qubits consisting of a linear array of four quantum dots with exchange coupling between nearest neighbors.  Other geometries can also be considered, but all will lead to a Hamiltonian of the same basic form. The exchange coupling between the middle two quantum dots serves as a controllable $X_1X_2$ coupling between the
two qubits, allowing for two-qubit gates to be performed\cite{LiPRB2012}. Unfortunately, the presence of this coupling interferes
with single qubit operations by introducing crosstalk between the two qubits. While this can be mitigated by simply reducing the
strength of the interaction, it can not be entirely eliminated, since the interaction cannot be completely turned off. In addition,
for future quantum computing applications with many qubits working together in a circuit, it is essentially impossible to turn
off all inter-qubit couplings while performing single qubit operations on specific qubits. This exchange coupling can also potentially
introduce leakage outside of the computational subspace.  Although the $z$-component of total spin is still conserved, the mixing
of spin states between qubits allows the system to enter, e.g., the state $\ket{\uparrow\uparrow\downarrow\downarrow}$, and conversely.
However, leakage can be guarded against by applying a large magnetic field difference between the two ST qubits\cite{LiPRB2012}.
We quantify exactly to what degree this protects against leakage by using perturbation theory.

Let us number the spins in one of our qubits 1 and 2, and those in the other qubit 3 and 4.  We now add the exchange
coupling, $J_{23}\vec{S}_2\cdot\vec{S}_3$, to the Hamiltonian for the full system, thus coupling the two qubits.  We now expand
this term out in terms of the $x$, $y$, and $z$ components of the spins.  The term $J_{23}S_{2z}S_{3z}$ simply allows
us to perform two-qubit operations; in terms of our computational basis, this term becomes $\tfrac{1}{4}J_{23}X_1X_2$.  Unfortunately,
this is also the term that results in crosstalk between the two qubits.  The terms involving the $x$ and $y$ components, on the other
hand, cause leakage out of the computational subspace; it is the effect of these terms that we now quantify.  In particular, we will
show that a large magnetic field difference between the two qubits helps to suppress leakage.

Let $B$ be the energy splitting of each leakage state due to the magnetic field difference between the two qubits. Denote the basis
states of the computational subspace by $\ket{n}$, where $n$ runs from 1 to 4, and denote the two leakage states by $\ket{\xi}$,
where $\xi$ runs over the two states $\ket{\uparrow\uparrow\downarrow\downarrow}$ and $\ket{\downarrow\downarrow\uparrow\uparrow}$. Let $H_4$ denote the terms of the Hamiltonian that connect states within the 4-dimensional
computational subspace, and, for simplicity, assume the basis $\ket{n}$ diagonalizes $H_4$, so $H_4\ket{n}=E_n\ket{n}$. We let the
unperturbed Hamiltonian be given by $H_4$ plus the terms corresponding to the energies of the leakage states, and we let the perturbation
be given by the terms that couple the computational states to the leakage states which we label $k_{n\xi}$ as follows:
\begin{equation}
H_0=H_4+B\ket{\uparrow\uparrow\downarrow\downarrow}\bra{\uparrow\uparrow\downarrow\downarrow}-B\ket{\downarrow\downarrow\uparrow\uparrow}\bra{\downarrow\downarrow\uparrow\uparrow}
\end{equation}
\begin{equation}
H'=\sum_{n,\xi}k_{n\xi}\ket{\xi}\bra{n}+k_{n\xi}^*\ket{n}\bra{\xi}
\end{equation}
The relative energy scales are such that $B\gg E_n\gg k_{n\xi}$.  Then using perturbation theory, the leading order corrections to
the eigenstates and energies are as follows:
\begin{equation}
\ket{n^1}=\sum_\xi\frac{\braket{\xi^0|H'|n^0}}{E_n^0-E_\xi^0}\ket{\xi^0}=\sum_\xi\frac{k_{n\xi}}{E_n\pm B}\ket{\xi^0}=O\bigg(\frac{k_{n\xi}}{B}\bigg)
\end{equation}
\begin{equation}
E_n^2=\sum_\xi\frac{|\braket{\xi^0|H'|n^0}|^2}{E_n^0-E_\xi^0}=\sum_\xi\frac{|k_{n\xi}|^2}{E_n\pm B}=O\bigg(\frac{k_{n\xi}^2}{B}\bigg)
\label{eqn:pertenergy}
\end{equation}
Thus, if a strong magnetic field difference is applied between the two singlet-triplet qubits, the effect of leakage due to exchange
coupling is of order $k/B$.  This is sufficiently small to ignore for applications of single-qubit gates, since $B$ can be made large
and the value of $k$ is already minimized to reduce crosstalk.

Since a sufficiently large magnetic field difference makes any terms which lead to leakage become negligible, we need only consider
the projection of the exchange term $J_{23}\vec{S}_2\cdot\vec{S}_3$ into the computational space.  For simplicity, we define $k=-J_{23}/4$, so that the interaction takes the form $kX_1X_2$.
Thus the effective Hamiltonian is given by
\begin{equation}
H_4=h_1X_1+h_2X_2+J_1Z_1+J_2Z_2+kX_1X_2.
\end{equation}

\section{First order expansions of crosstalk and noise}

We closely follow the technique developed in Ref. \onlinecite{ButerakosPRB2018}, to perform an expansion of the evolution operator $e^{-itH}$ to first
order in $k$. This expansion uses commutation relations of Pauli matrices in order to separate the single-qubit parts of the evolution
operator, $e^{-it(h_iX_i+J_iZ_i)}$, from the two-qubit cross terms. To simplify the resulting expressions, we use the shorthand $a_i=\sqrt{h_i^2+J_i^2}$,
and introduce the following rotated basis, chosen so that the single qubit evolution for qubit $i$ reduces to an $X_i'$ rotation.
\begin{align}
\begin{split}
&X_i'=(h_iX_i+J_iZ_i)/a_i\\
&Y_i'=Y_i\\
&Z_i'=(-J_iX_i+h_iZ_i)/a_i
\label{eqn:paulirotation}
\end{split}
\end{align}
This expansion allows the evolution operator to be written in the form
\begin{equation}
e^{-itH}=e^{-it(a_1X_1'+a_2X_2')}\left (1-i\sum_{i'j'}\Delta_{i'j'}^{\text{ex}}\sigma_{i'}\otimes\sigma_{j'}\right ),
\label{eqn:concisekerror}
\end{equation}
where $\sigma_{i'}$ runs over the rotated Pauli matrices $X_1'$, $Y_1'$, $Z_1'$, and similarly $\sigma_{j'}$ runs over $X_2'$, $Y_2'$,
$Z_2'$.  Performing the expansion yields the coefficients $\Delta_{i'j'}^{\text{ex}}$ as follows:
\begin{align}
\Delta_{X_1'X_2'}^{\text{ex}}=k&\frac{h_1h_2}{a_1a_2}t,\nonumber\\
\begin{split}
\Delta_{Y_1'Y_2'}^{\text{ex}}=k&\frac{J_1J_2}{2a_1a_2}\Big\{-t\sinc{[2(a_1+a_2)t]}\\&+t\sinc{[2(a_1-a_2)t]}\Big\},
\end{split}\nonumber\\
\begin{split}
\Delta_{Z_1'Z_2'}^{\text{ex}}=k&\frac{J_1J_2}{2a_1a_2}\Big\{t\sinc{[2(a_1+a_2)t]}\\&+t\sinc{[2(a_1-a_2)t]}\Big\},
\end{split}\nonumber\\
\Delta_{X_1'Y_2'}^{\text{ex}}=-&k\frac{h_1J_2}{a_1}t^2\sinc^2{(a_2t)},\nonumber\\
\Delta_{X_1'Z_2'}^{\text{ex}}=-&k\frac{h_1J_2}{a_1a_2}t\sinc{(2a_2t)},\nonumber\\
\begin{split}
\Delta_{Y_1'Z_2'}^{\text{ex}}=k&\frac{J_1J_2}{2a_1a_2}\Big\{(a_1+a_2)t^2\sinc^2{[(a_1+a_2)t]}\\&+(a_1-a_2)t^2\sinc^2{[(a_1-a_2)t]}\Big\}.
\end{split}
\label{eqn:kerror}
\end{align}
The other three terms ($\Delta_{Y_1'X_2'}^{\text{ex}}$, $\Delta_{Z_1'X_2'}^{\text{ex}}$, and $\Delta_{Z_1'Y_2'}^{\text{ex}}$) are omitted
for the sake of brevity, but can be obtained by interchanging the subscripts 1 and 2 in the expressions above. Since $J_i$ is time dependent,
the rotated matrices $X_i'$, $Y_i'$, $Z_i'$ are also time dependent, and thus are transformed back into the standard basis before proceeding
further into the calculation.

In addition to crosstalk, the singlet-triplet qubit system encounters noise from two separate sources, namely, field and charge noise. We
work in the quasistatic approximation, where we assume that the noise changes slowly compared to the gate implementation time.  These noise sources enter into the
Hamiltonian via corrections to the values $h_i$ and $J_i$. Since the magnetic field itself does not vary over the implementation of the gate,
the field noise simply causes a small constant term $\textit{dh}_i$ to be added to $h_i$. Charge noise, however, is more subtle,
since the value of $J_i$ is changed over the course of the pulse in order to implement each specific gate. If the exchange interaction is adjusted via detuning control, that is, by changing the energy difference $\epsilon_i$ between the two dots comprising a qubit, then the change in $J_i$ is given by $\textit{dJ}_i=\frac{\partial J_i}{\partial\epsilon_i}\,d\epsilon_i$, where $d\epsilon_i$ is caused by the charge noise and stays constant in the quasistatic limit.
Empirically, $J_i$ is found to have an exponential dependence on $\epsilon_i$, at least in the regime
of operation, which means that $\textit{dJ}_i$ is proportional to $J_i$. It is possible that some systems could have a different dependence of
$J_i$ on $\epsilon_i$, but it has been shown that \textsc{supcode} can easily accommodate these other cases\cite{WangPRA2014}.  For specificity,
we consider the case of $dJ_i$ being proportional to $J_i$, i.e., $J_i$ being exponential in $\epsilon_i$.

The expansion of the evolution operator for a single qubit to first order in noise terms has been performed\cite{KestnerPRL2013}, and since the noise
sources as well as the coupling between qubits are small, any effect of one on the other can be ignored. Thus we simply combine the crosstalk
expansion terms we obtained with the previously derived noise terms $\Delta_i^{\text{q}n}$, so that, to first order, the full noisy evolution operator
is given by
\begin{align}
\begin{split}
&e^{-itH}=e^{-it(h_1X_1+J_1Z_1)}e^{-it(h_2X_2+J_2Z_2)}\times\\ &\bigg[1-i\sum_i(\Delta_i^{\text{q1}}\sigma_i\otimes 1
+\Delta_i^{\text{q2}}1\otimes\sigma_i) -i\sum_{ij}\Delta_{ij}^{\text{ex}}\sigma_i\otimes\sigma_j\bigg]
\label{eqn:evolutionoperator}
\end{split}
\end{align}
with $\Delta_i^{\text{q}n}$ given by
\begin{align}
\begin{split}
\Delta_x^{\text{q}n}=&\frac{2h_n^2a_nt+J_n^2\sin{2a_nt}}{2a_n^3} \mathit{dh}_n\\&+\frac{h_nJ_n(2a_nt-\sin{2a_nt})}{2a_n^3}\mathit{dJ}_n,
\end{split}\nonumber\\
\begin{split}
\Delta_y^{\text{q}n}=&\frac{J_n(\cos{2a_nt}-1)}{2a_n^2} \mathit{dh}_n\\&+\frac{h_n(1-\cos{2a_nt})}{2a_n^2}\mathit{dJ}_n,
\end{split}\nonumber\\
\begin{split}
\Delta_z^{\text{q}n}=&\frac{h_nJ_n(2a_nt-\sin{2a_nt})}{2a_n^3} \mathit{dh}_n\\&+\frac{2J_n^2a_nt+h_n^2\sin{2a_nt}}{2a_n^3}\mathit{dJ}_n.
\end{split}
\label{eqn:singleerror}
\end{align}
The evolution operator given in Eq.\ \eqref{eqn:evolutionoperator}, is of the form $U=R(1+\Delta)$, where $R$ is an ideal rotation of qubits 1
and 2 along particular axes of the Bloch sphere, and $\Delta$ is some small error produced by noise and crosstalk. Uncorrected rotations along
axes in the first quadrant of the $xz$-plane can be performed simply by allowing the system to evolve at a fixed value of $J_1$ and $J_2$ for a
given amount of time $t$. In order to perform rotations about other axes or to correct for crosstalk, it is necessary to perform several uncorrected
rotations in a row (i.e., to allow the system to evolve for some time $t_1$, then change the values of $J_1$ and $J_2$ and allow the system to
continue evolving for some time $t_2$, etc.). In this case, it is necessary to calculate the errors for each segment of the gate separately and
combine the results, since the values of $J_1$ and $J_2$ are different for each part. To first order, the result of performing an
uncorrected gate $U_2$ followed by $U_1$ is
\begin{equation}
U_1U_2=R_1(1+\Delta_1)R_2(1+\Delta_2)=R_1R_2(1+R_2^\dagger\Delta_1R_2+\Delta_2).
\label{eqn:adderrors}
\end{equation}
For the pulses we generate, we will set $h_1=h_2=h$; however, the same method can be applied to systems where $h_1$ and $h_2$ are different. In
either case, the noise terms $\mathit{dh}_1$ and $\mathit{dh}_2$ are independent.

\section{Error Cancellation}

As with previous \textsc{supcode} pulses, our strategy involves applying an initial uncorrected gate followed by an uncorrected identity operation
in such a way that the errors cancel out to leading order. Single qubit rotations have been studied extensively in Ref.\ \onlinecite{ThrockmortonPRB2017},
so for the initial uncorrected gate, we use the equations presented in that work to generate the desired rotation on qubit 1, and perform a $2\pi$
rotation on qubit 2 at the same time. The uncorrected identity operation which follows can be somewhat complex, and in fact, a sufficient degree
of complexity is required in order for it to have the freedom to cancel all sources of error for an arbitrary gate. The general form of this identity
operation is the key to generating fast, efficient pulse sequences. For a single-qubit system\cite{WangPRA2014}, a sequence of interrupted $2\pi$
rotations was used, where the axes of rotation were tuned to precisely cancel out the error. Specifically, the total error of the pulse was calculated
in terms of parameters $j_n$ corresponding to the values of $J$ at each part of the gate, and then the total error was set equal to zero, forming
a complicated set of equations which were solved numerically for the parameters $j_n$. Correcting a two-qubit system is significantly more complicated
because the amount of time needed to perform $2\pi$ rotations differs depending on the axis of rotation, and if qubits 1 and 2 are rotated about different axes concurrently, one operation will finish before the other.  Freedom to vary the axes of rotation is needed to allow for numerical solutions which eliminate error, but
varying these values can affect the timing between qubits 1 and 2, specifically which segment of the pulse on qubit 1 coincides with a given segment on
qubit 2. This makes expressing the total error as a function of a set of parameters, a key step in the error correction process, very difficult.

In order to avoid this problem, Ref.\ \onlinecite{ButerakosPRB2018} uses a base identity operation consisting of two $2\pi$ rotations on one qubit,
and a $4\pi$ rotation on the other, chosen such that the time taken to perform the $4\pi$ rotation equals the total time needed to perform both
$2\pi$ rotations. While this form of an identity operation solves the problem of timing, it can be somewhat constricting, leading to longer, slower
pulses. As stated previously, the major difficulty to performing dynamical decoupling on singlet-triplet qubit systems is the restriction of rotations
to the first quadrant of the $xz$-plane. However, in order to generate efficient pulses, it is necessary to have a large amount of freedom in rotating
qubits during the pulse. For singlet-triplet systems, this is best accomplished by allowing sequences with values of $J_i$ alternating between large
and small. The scheme that uses a $4\pi$ rotation does not allow for strictly alternating values of $J_i$, since the basic pattern was a nested
sequence of $2\pi$, $2\pi$, $4\pi$ rotations, corresponding to a cycle between small, large, and medium $J_i$.  Additionally, the value of $J_i$
during the $4\pi$ rotation is completely constrained by the parameters chosen for the opposite qubit, meaning that the qubits cannot be individually
controlled.

\begin{figure}
	\includegraphics[width=\columnwidth]{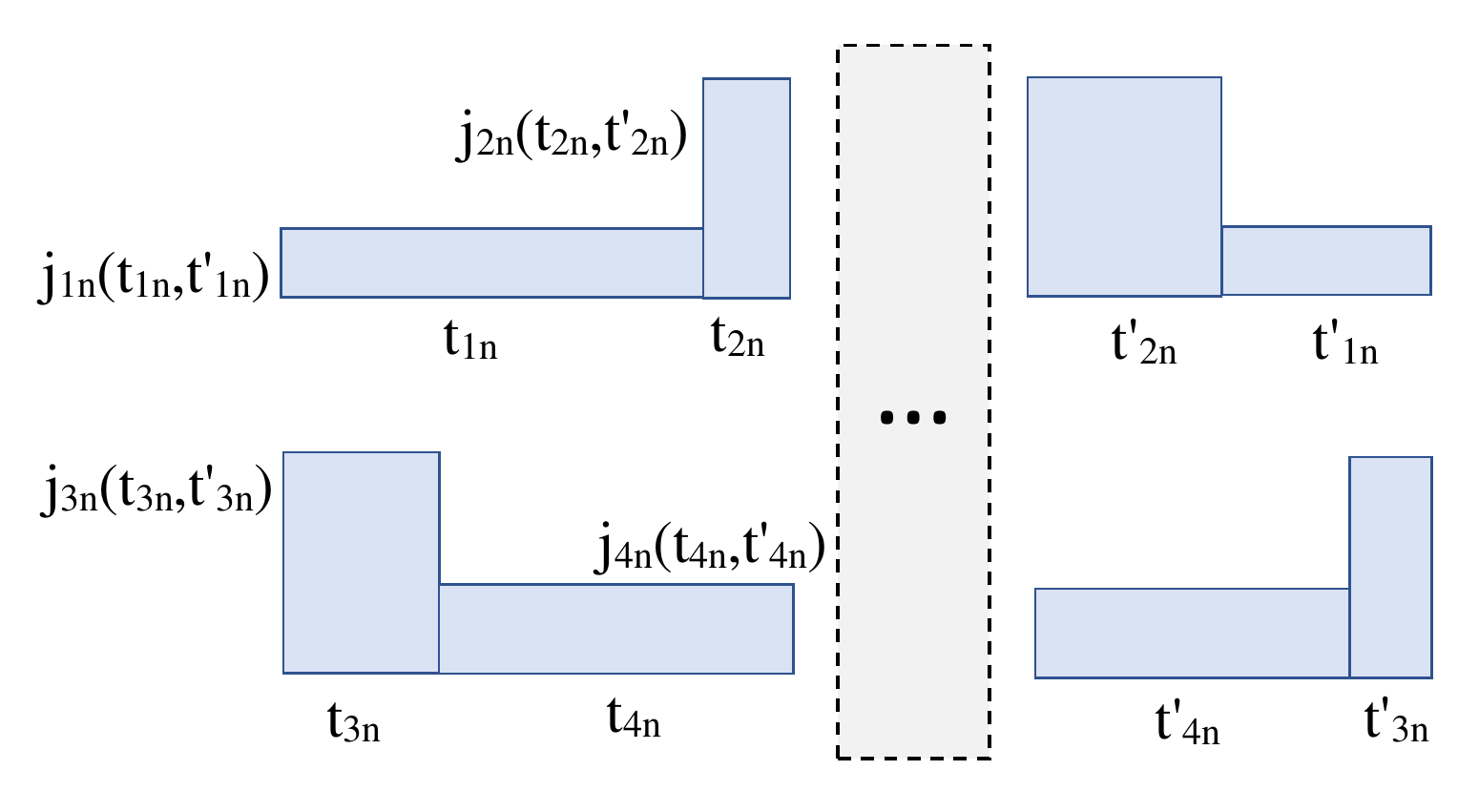}
	\caption{The base uncorrected identity which consists of two $2\pi$ rotations on each qubit (top and bottom). $t_{1n}$ and $t_{1n}'$ define one
	$2\pi$ rotation on qubit 1, and this rotation is interrupted by the $2\pi$ rotation given by $t_{2n}$ and $t_{2n}'$. Similarly, $t_{3n}$, $t_{3n}'$,
	$t_{4n}$, and $t_{4n}'$ define two $2\pi$ rotations on qubit 2. The ellipsis in the middle indicates where copies of this base identity with
	larger values of $n$ interrupt this identity.}
	\label{fig:improvedpulse}
\end{figure}

We are able to improve upon the previous method by replacing the $4\pi$ rotation with two $2\pi$ rotations, while still enforcing the constraint that
the total time of these two rotations must equal the total time of the two $2\pi$ rotations on the other qubit, as shown in the figure (\ref{fig:improvedpulse}).
This allows an alternating pattern between large and small values of $J_i$, and reduces the dependence of one qubit's control on the other. To simplify
some of the resulting equations, we use the times of each segment of the pulse as the free parameters, rather than the pulse height $j_i$ used in previous
works. Ultimately these are equivalent, since for a fixed value of $h$, the strength of the exchange interaction $J_i$ forms a one-to-one correspondence
with the time needed to perform a $2\pi$ rotation, as discussed below. We define times $t_{1n}$ through $t_{4n}$ and $t_{1n}'$ through $t_{4n}'$ as in
Fig.\ \ref{fig:improvedpulse}, where the segments of the pulse corresponding to $t_{in}'$ and $t_{in}'$ form a single interrupted $2\pi$ rotation. The
subscript $n$ distinguishes between the different nested copies of the base identity operation. We impose the constraint that $t_{1n}+t_{2n}=t_{3n}+t_{4n}$,
and similarly for $t'$, meaning that each copy of the base identity has 6 independent parameters. The values of $j_{in}$ are determined in terms of the total
time $t_{in}+t_{in}'$ by calculating what value of $j_{in}$ gives a $2\pi$ rotation which takes the given amount of time, as follows:
\begin{equation}
j_{in}=\sqrt{\Big(\frac{\pi}{t_{in}+t_{in}'}\Big)^2-h^2}.
\label{eqn:jprime}
\end{equation}

To write the full sequence of operations comprising the nested identity operation, let $U(J_1,J_2,t)$ be an uncorrected rotation at the given values $J_1$,
and $J_2$, for a time $t$. The sequence of operations before and after the interruption, shown on the left and right of Fig.\ \ref{fig:improvedpulse}, we
denote as $A_n$ and $A_n'$ respectively. These each consist of three rotations: the first and third corresponding to the first and second $2\pi$ rotations
on each qubit, and the middle corresponding to the overlap region where part of the first $2\pi$ rotation on one qubit is being performed concurrently with
the second $2\pi$ rotation on the other. This overlap requires a selection statement dependent on which $2\pi$ rotations are overlapping, which is determined
by the times $t_{1n}$ and $t_{3n}$. The sequences $A_n$ and $A_n'$ are given by
\begin{equation}
A_n=
\begin{cases}
U(j_{2n},j_{4n},t_{2n})U(j_{1n},j_{4n},t_{1n}-t_{3n})\\
\qquad\qquad\qquad\times U(j_{1n},j_{3n},t_{3n})\text{\qquad\quad if $t_{1n}>t_{3n}$,}\\
U(j_{2n},j_{4n},t_{4n})U(j_{2n},j_{3n},t_{3n}-t_{1n})\\
\qquad\qquad\qquad\times U(j_{1n},j_{3n},t_{1n})\text{\qquad\quad if $t_{1n}<t_{3n},$}
\end{cases}
\label{eqn:idtleft}
\end{equation}
\begin{equation}
A_n'=
\begin{cases}
U(j_{1n},j_{3n},t_{3n}')U(j_{1n},j_{4n},t_{1n}'-t_{3n}')\\
\qquad\qquad\qquad\times U(j_{2n},j_{4n},t_{2n}')\text{\qquad\quad if $t_{1n}'>t_{3n}'$,}\\
U(j_{1n},j_{3n},t_{1n}')U(j_{2n},j_{3n},t_{3n}'-t_{1n}')\\
\qquad\qquad\qquad\times U(j_{2n},j_{4n},t_{4n}')\text{\qquad\quad if $t_{1n}'<t_{3n}'$.}
\end{cases}
\label{eqn:idtright}
\end{equation}
We nest multiple copies of the base identity operation in order to create enough degrees of freedom to find a solution that cancels the error of the initial rotation.
We find that 5 is the minimum number of copies needed, and so the sequence of rotations which comprise the full uncorrected identity, denoted $I^{(5)}$, is as follows:
\begin{equation}
I^{(5)}=\prod_{n=1}^5A_n'\prod_{n=5}^1A_n.
\label{eqn:identityfull}
\end{equation}
The error for each uncorrected rotation is calculated in terms of the parameters $t_{in}$ and $t_{in}'$ by Eq.\ \eqref{eqn:evolutionoperator}, and added
together using Eq.\ \eqref{eqn:adderrors}, and the result is added to the error of the initial pulse. The norm of this total first order error is numerically
minimized over the parameters $t_{in}$ and $t_{in}'$, and a minimum sufficiently close to zero indicates that the crosstalk and static noise error has been
canceled using the pulse sequence given by the values of $t_{in}$ and $t_{in}'$. Experimental constraints on the value of $J_i$ can be accounted for by defining
bounds on the total time of a $2\pi$ rotation $t_\text{min/max}$ in terms of the constraining values of $J_\text{min/max}$:
\begin{equation}
t_\text{min/max}=\frac{\pi}{\sqrt{h^2+J_\text{max/min}^2}}.
\end{equation}
Then, during the numerical minimization, the constraint on the time for each $2\pi$ rotation $t_\text{min}<t_{in}+t_{in}'<t_\text{max}$ is applied, along with
the physical constraint that $t_{in},t_{in}'>0$. This ensures that the derived pulse sequence respects experimental limitations. For the pulse sequences we
derived, we used values of $J_\text{min}$ and $J_\text{max}$ equal to $\frac{1}{30}$ and $30$ respectively, but other constraints can be used in much the same
manner.

\section{Numerical Results}
\begin{figure}
	\includegraphics[width=\columnwidth]{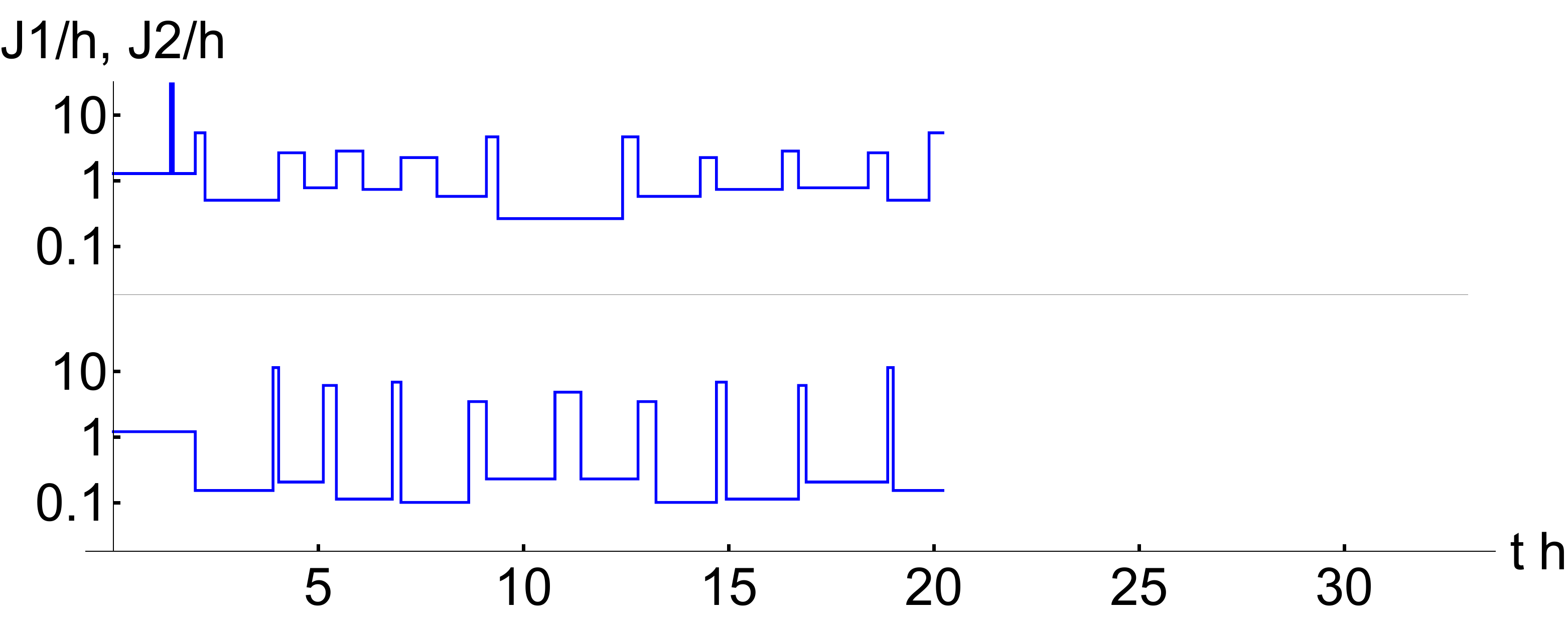}
	\includegraphics[width=\columnwidth]{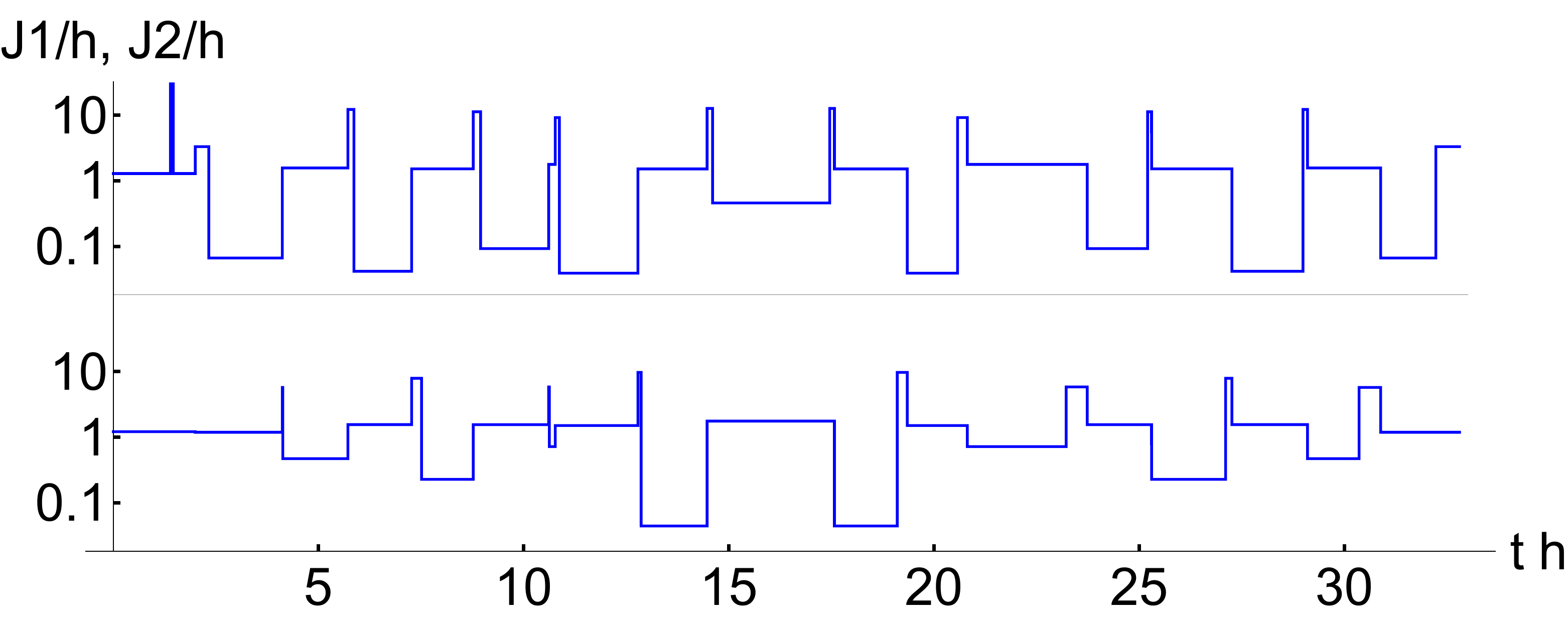}
	\caption{{\bf Top:} A corrected pulse sequence for the gate $e^{-\frac{2\pi}{3}\frac{X+Y+Z}{\sqrt{3}}}$ using the identity operation presented in this work.
	{\bf Bottom:} For comparison, a corrected pulse sequence for the same gate using the identity with $4\pi$ rotations.}
	\label{fig:newmethod}
\end{figure}

We generated pulses which correct against first order error for the 24 Clifford gates. The parameters precisely defining these pulses are given in the Appendix.
For comparison, we also generated pulses using the previous method which uses $4\pi$ rotations. Using the identity operation with two $2\pi$ rotations, we are
able to reduce the total length of the error correcting identity from $36\pi$ to $20\pi$, corresponding to roughly a 40\% decrease in the length of the pulse, as
shown in figure \ref{fig:newmethod}.

\begin{figure}[b]
	\includegraphics[width=\columnwidth]{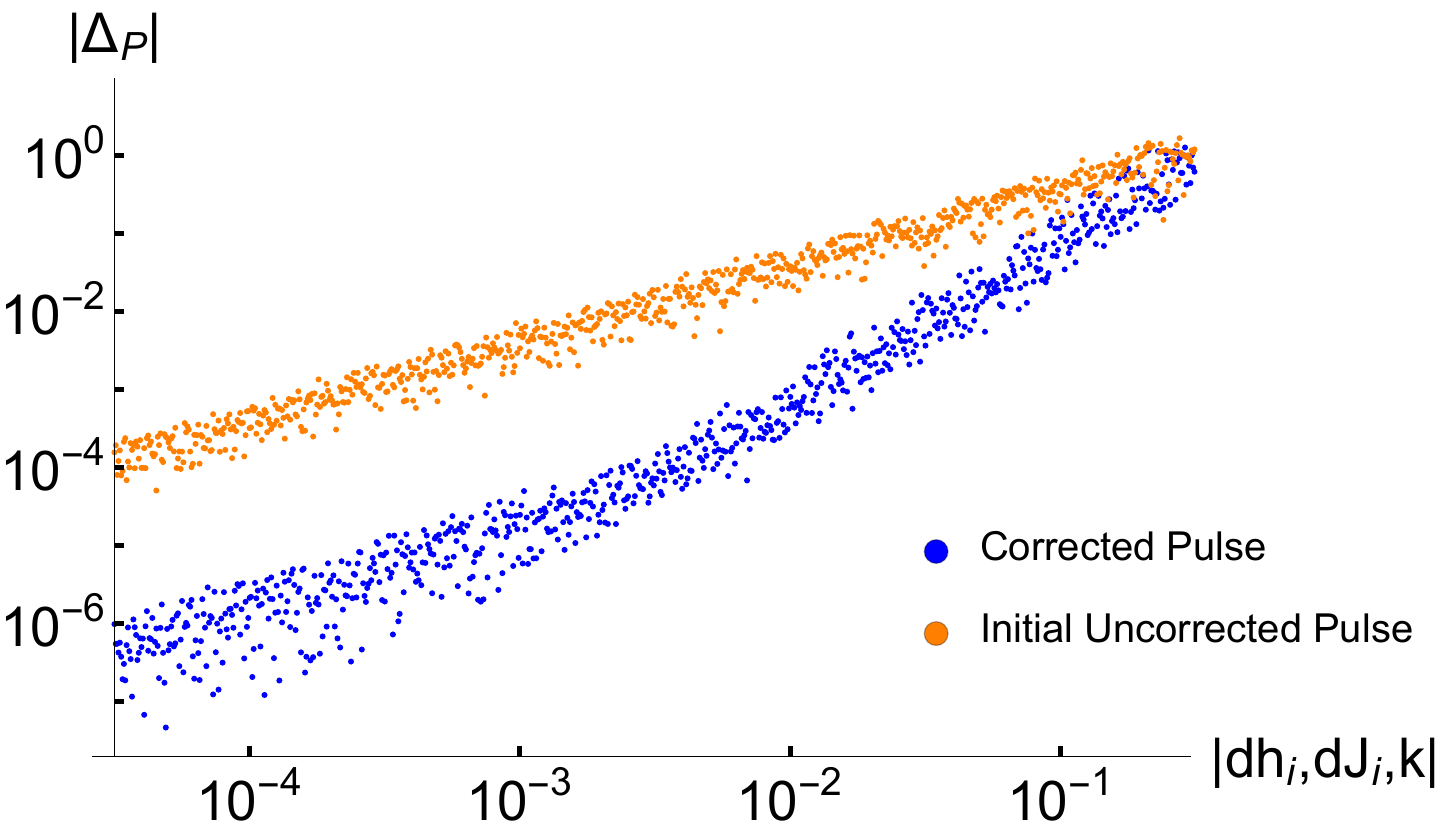}
	\includegraphics[width=\columnwidth]{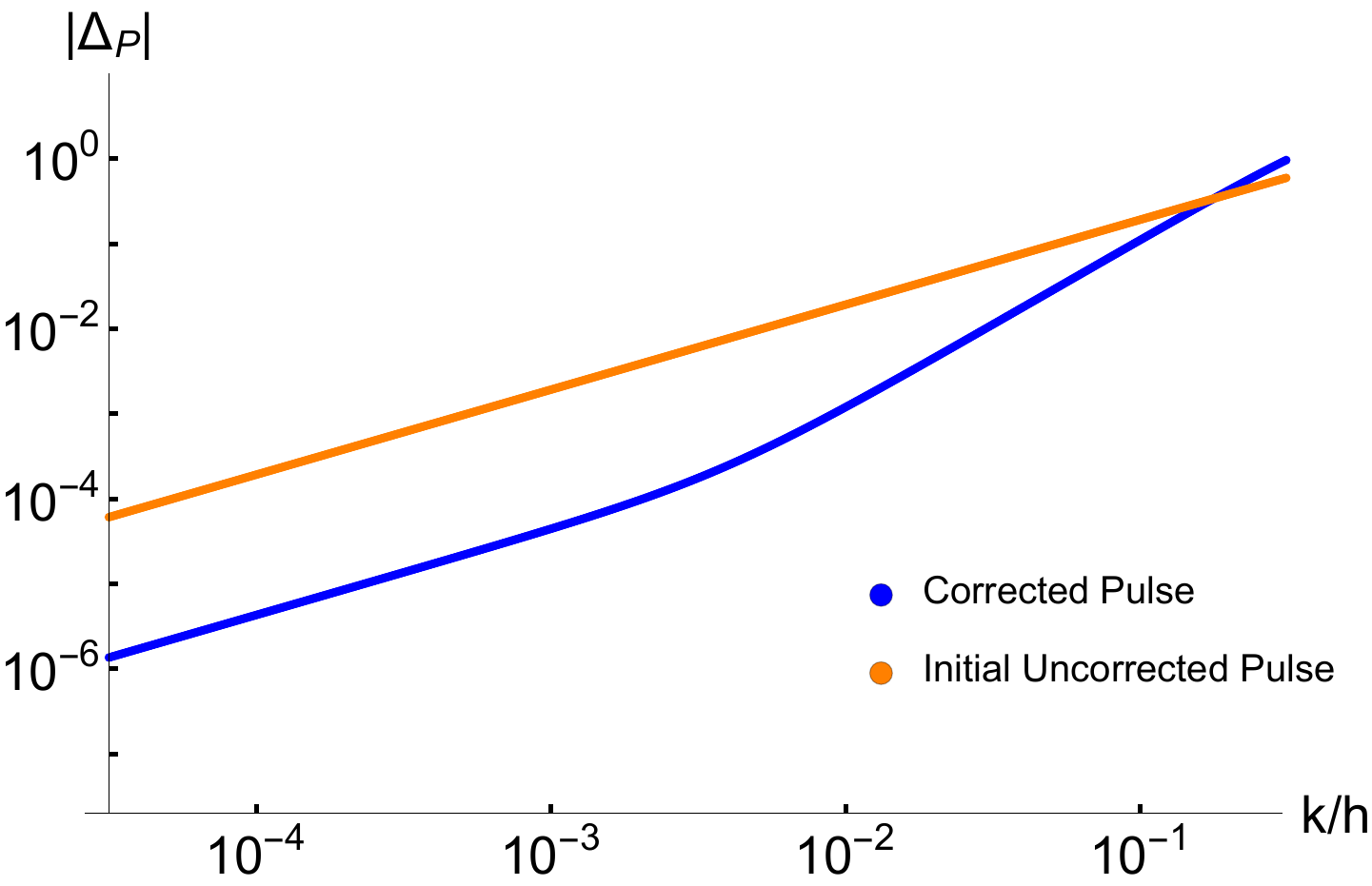}
	\caption{Error scaling of a naive uncorrected rotation compared with the corrected pulse sequence for the gate $e^{-\frac{2\pi}{3}\frac{X+Y+Z}{\sqrt{3}}}$ at
	a constant value of $B=100h$. {\bf Top:} Error with random values of $(\mathit{dh}_i/h, \mathit{dJ}_i/J_i, k/h)$. {\bf Bottom:} Error plotted against only $k/h$, with $\mathit{dh}_i=\mathit{dJ}_i=0$.}
	\label{fig:errorcheck}
\end{figure}

A full dynamical decoupling randomized benchmarking analysis is needed to test the pulses against experimental noise.  Such an analysis, while feasible, is extremely
computationally demanding, and therefore should only be carried out in the context of an actual experimental realization of the pulse sequence developed in the current
work.  However, we can demonstrate that under a quasistatic noise approximation, the sequences we derived completely cancel first order error. This is done by choosing
random values for the errors $\mathit{dh}_i/h$, $\mathit{dJ}_i/J_i$ and the relative strength of the exchange coupling $k/h$. From these, the uncorrected rotations
performed for each segment of the pulse are evaluated by numerically exponentiating the full 6-dimensional Hamiltonian which includes the noise terms, leakage states,
and magnetic field splitting $B$. By multiplying together the matrices corresponding to each individual segment, the matrix resulting from applying the full pulse with the given noise terms can be found. We denote this matrix $P(\mathit{dh}_i, \mathit{dJ}_i, k)$ and compare it against the ideal Clifford gate which the pulse implements, obtaining a difference $\Delta_P$. The norm of $\Delta_P$, defined as $\sqrt{\text{Tr}(\Delta_P^\dagger\Delta_P)}$,
is related to the infidelity of the pulse $P$ averaged over all initial states, and thus is a good measure of total error of the pulse. In order to determine how the error of the pulse scales with initial noise terms, the norm of $\Delta_P$ is plotted against the norm of $(\mathit{dh}_i/h,\mathit{dJ}_i/J_i,k/h)$.
In the top portion of Fig.\ \ref{fig:errorcheck}, we show such a plot
for a value of $B$ equal to $100h$. As shown, the total error is second order for initial error values greater than roughly $10^{-2}$, but becomes linear below that point
as leakage errors begin to dominate. Increasing or decreasing the value of $B$ will cause this crossover point to shift left or right respectively.
One can isolate the behavior of the error in terms of the interaction strength $k$ alone by performing a similar calculation setting $\mathit{dh}_i$ and $\mathit{dJ}_i$ to 0. This shows the error scaling of the corrected and uncorrected pulses more clearly. Specifically, from the bottom of Fig.\ \ref{fig:errorcheck}, we again see that the error scales quadratically above $10^{-2}$ and linearly below that point, yielding an improvement of nearly two orders of magnitude. The reason for the spread in data points in the top graph compared to the bottom, is that for randomly selected noise values, $(\mathit{dh}_i/h, \mathit{dJ}_i/J_i)$ it is possible for different second order error terms to add constructively or destructively, thus adding some variation to the total error of the pulse. Restricting to only one variable produces a much finer line, as this is no longer the case. In both cases, we see the elimination of first order error for noise values above $1/B$.

\section{Conclusion}

We have demonstrated a method for correcting crosstalk- and noise-induced error in single-qubit gates in exchange-coupled singlet-triplet qubits.  Unlike the capacitively-coupled
qubits studied in Ref.\ \onlinecite{ButerakosPRB2018}, we can tune the coupling between the qubits independently of the intraqubit exchange coupling used to perform gates.  However,
leakage out of the computational subspace is a problem here, which does not arise for capacitive coupling.  We first showed that a sufficiently large magnetic field difference
between the two qubits helps to suppress leakage, and then proceeded to develop pulse sequences that cancel crosstalk- and noise-induced error to first order.  Our methods are
similar to those used in Ref.\ \onlinecite{ButerakosPRB2018} for capacitively coupled singlet-triplet qubits---we perform an uncorrected gate, and then follow it with a sequence
of uncorrected identity operations designed in such a way as to cancel crosstalk- and noise-induced error in the gate to first order.  The basic building blocks of our sequences,
however, are different---rather than perform a single $4\pi$ rotation on the ``idle'' qubit, we perform two $2\pi$ rotations, each with a different value of the intraqubit exchange
coupling.  This allows for shorter and faster sequences; we now only require at least 5 of these ``blocks'', fewer than what was needed with the older method.

We find that our sequences do, in fact, reduce the error in our qubits by several orders of magnitude.  We notice, however, that, for low noise and crosstalk, the error is still first
order in our measure of the total error described above, but then crosses over to second-order behavior for larger values.  This indicates that, compared to the analogous sequences for
capacitively-coupled qubits, our ability to use the sequences presented in this work to combat error is more limited.  While, as usual, our results show that reduction of noise and
crosstalk during a single-qubit gate are important for improving fidelity, we also show that creating a large magnetic field difference between the two qubits will help to achieve
this goal in the case of exchange-coupled qubits.  While a number of techniques exist for reducing magnetic field noise, such as polarization of nuclear spins in GaAs or working with
isotopically-purified Si, which has very few magnetic impurities, charge noise is much more difficult to handle, in no small part due to the fact that the origin of this noise is poorly
understood.

We should also point out that, throughout this work, we made a number of approximations.  First of all, we assumed that the magnetic field gradients on the two qubits were the same, which
we believe is a reasonable assumption to make.  In principle, they could differ, either intentionally or due to natural variation in the field gradients produced at each qubit, whatever the
method used to do so might be.  Our method already corrects for small, unintentional, variations, since these can essentially be included as part of the noise term.  Larger variations, however,
would require us to modify our pulse sequences.  This also introduces complications similar to those in the capacitively-coupled case\cite{ButerakosPRB2018}.  Second of all, we assumed that
the pulses were perfect square pulses.  Such pulses are impossible in reality, as there will be a finite rise or fall time.  The effect of this finite ramping up or down has been studied in
the case of a single isolated qubit\cite{WangPRA2014}, and it was found that the sequences derived therein for correcting noise-induced error still performed well.  We thus expect that this
will continue to hold true even for the sequences derived in this work.  Finally, we assumed that the noise in the system was quasistatic, i.e., it was possible to neglect the time variation
in the noise.  While this tends to be a good approximation, it is found that, in reality, both types of noise in singlet-triplet qubits exhibit power-law spectra.  To be exact, both types
of noise follow a $1/f^\alpha$ spectrum, with $\alpha=2.6$ for the magnetic field noise and $0.7$ for the charge noise\cite{DialPRL2013,MedfordPRL2012}.  Filtering out the high-frequency
noise components, especially in the charge noise, will bring a system closer to the quasistatic limit that we worked in.  It has been shown, however, that, even in this case, sequences
similar to ours for a single isolated qubit\cite{WangPRA2014,ThrockmortonPRB2017} still correct errors due to time-dependent noise, and thus we expect the same to hold true for the sequences
developed here.  Developing a means of combatting noise that takes into full account the time dependence of any specific noise spectrum will likely yield even better results, but doing so is
beyond the scope of this work.

Even though we considered the case of two coupled qubits here, an eventual practical quantum computer will have billions of qubits, all with some degree of coupling.  As a result,
extensions of the techniques developed here to the case of a larger number of qubits will be necessary.  Since the coupling between two qubits falls off with distance, we expect that one really
only needs to correct for error due to qubits up to a distance of, say, the fifth-nearest neighbors, if that.  Correction of errors due to qubits further away should be possible through other
error-correction techniques.

\acknowledgments
This work is supported by the Laboratory for Physical Sciences.

\appendix
\section{Parameters for Dynamical Decoupling Pulse Sequences for the 24 Clifford Gates}

In Tables \ref{Tab:RotParamsI}--\ref{Tab:RotParamsIII} we list parameters defining pulse sequences for the 24 Clifford gates. The top portion of the tables consists of values $j_i^{\text{rot}}$,
$t_i^{\text{rot}}$, $j'^{\text{rot}}$ which encode the initial rotations as
\begin{equation}
R=\prod_{i=1}^N U(j_i^{\text{rot}},j'^{\text{rot}},t_i^{\text{rot}})
\end{equation}
for $N$ equal to 1, 3, or 5, depending on the number of values given in the table. The rest of the tables are the values of $t_{in}$ and $t'_{in}$ which define the uncorrected
identity by Eqs.\ \eqref{eqn:jprime}--\eqref{eqn:identityfull}. Only values of $i$ from 1 to 3 are shown with $t_{4n}$ being defined as $t_{1n}+t_{2n}-t_{3n}$, and similarly
for $t_{4n}'$.

\begin{table*}
	\centering
\begin{tabular}{|c|c|c|c|c|c|c|c|c|}
\hline
Axis & $I$ & $x$ & $x$ & $x$ & $y$ & $y$ & $y$ & $z$ \\
\hline
Angle & 0 & $\pi/2$ & $\pi$ & $3\pi/2$ & $\pi/2$ & $\pi$ & $3\pi/2$ & $\pi/2$ \\
\hline
$j'^{\text{rot}}$ & 30.0000 & 3.37447 & 1.57530 & 0.80106 & 0.82357 & 0.80164 & 0.77982 & 6.50963 \\
\hline
$j_1^{\text{rot}}$ & 30.0000 & 0.19724 & 0.19336 & 0.20358 & 30.0000 & 30.0000 & 30.0000 & 30.0000 \\
\hline
$j_2^{\text{rot}}$ & --- & 30.0000 & 30.0000 & 30.0000 & 0.93548 & 0.93548 & 0.93548 & 6.54409 \\
\hline
$j_3^{\text{rot}}$ & --- & 0.19724 & 0.19336 & 0.20358 & 30.0000 & 30.0000 & 30.0000 & 30.0000 \\
\hline
$j_4^{\text{rot}}$ & --- & --- & --- & --- & 0.93548 & 0.93548 & 0.93548 & --- \\
\hline
$j_5^{\text{rot}}$ & --- & --- & --- & --- & 30.0000 & 30.0000 & 30.0000 & --- \\
\hline
$t_1^{\text{rot}}$ & 0.10466 & 0.39633 & 0.79278 & 1.17605 & 0.02617 & 0.02617 & 0.02617 & 0.01640 \\
\hline
$t_2^{\text{rot}}$ & --- & 0.09997 & 0.09813 & 0.09981 & 1.14711 & 1.14711 & 1.14711 & 0.44421 \\
\hline
$t_3^{\text{rot}}$ & --- & 0.39633 & 0.79278 & 1.17605 & 0.02617 & 0.05233 & 0.07850 & 0.01640 \\
\hline
$t_4^{\text{rot}}$ & --- & --- & --- & --- & 1.14711 & 1.14711 & 1.14711 & --- \\
\hline
$t_5^{\text{rot}}$ & --- & --- & --- & --- & 0.07850 & 0.07850 & 0.07850 & --- \\
\hline
$t_{1,1}$ & 0.23932 & 0.52518 & 0.17782 & 0.35794 & 0.87280 & 1.40951 & 1.47408 & 2.08110 \\
\hline
$t_{2,1}$ & 1.09125 & 1.69460 & 2.08781 & 1.29553 & 0.18612 & 0.20762 & 0.35782 & 0.22124 \\
\hline
$t_{3,1}$ & 1.12263 & 1.71424 & 1.94402 & 1.20225 & 0.24035 & 0.28494 & 0.23108 & 0.56868 \\
\hline
$t'_{1,1}$ & 0.59780 & 0.35620 & 0.40442 & 0.25259 & 2.16299 & 1.40150 & 1.30849 & 1.04347 \\
\hline
$t'_{2,1}$ & 1.61793 & 1.40675 & 1.05123 & 1.31383 & 0.21544 & 0.48157 & 0.45315 & 0.18968 \\
\hline
$t'_{3,1}$ & 1.98543 & 1.25467 & 0.69678 & 1.26393 & 0.41602 & 0.17172 & 0.28890 & 0.27459 \\
\hline
$t_{1,2}$ & 0.44412 & 0.30496 & 0.21167 & 0.33209 & 1.80820 & 1.28984 & 1.17247 & 1.28622 \\
\hline
$t_{2,2}$ & 1.43763 & 1.30979 & 1.62248 & 1.40209 & 0.23931 & 0.44525 & 0.33085 & 0.24893 \\
\hline
$t_{3,2}$ & 1.58580 & 1.20774 & 1.51527 & 1.14195 & 0.61849 & 0.19688 & 0.27989 & 0.39266 \\
\hline
$t'_{1,2}$ & 0.43189 & 0.32979 & 0.22604 & 0.44226 & 1.32490 & 1.43301 & 1.61739 & 1.81344 \\
\hline
$t'_{2,2}$ & 1.33166 & 1.33003 & 1.32999 & 1.68187 & 0.25344 & 0.24189 & 0.17401 & 0.24925 \\
\hline
$t'_{3,2}$ & 1.49696 & 1.20594 & 1.39791 & 1.56451 & 0.26449 & 0.19837 & 0.22329 & 0.46107 \\
\hline
$t_{1,3}$ & 0.59437 & 0.19836 & 0.32108 & 0.23737 & 1.63651 & 1.90912 & 0.78061 & 1.32820 \\
\hline
$t_{2,3}$ & 1.27812 & 1.69950 & 1.08601 & 1.62730 & 0.31245 & 0.20040 & 0.99287 & 0.21810 \\
\hline
$t_{3,3}$ & 1.71539 & 1.47222 & 0.47707 & 1.62737 & 0.53524 & 0.39193 & 0.31686 & 0.81853 \\
\hline
$t'_{1,3}$ & 0.83936 & 0.20294 & 0.25174 & 0.24861 & 1.46388 & 0.77040 & 0.87515 & 1.69784 \\
\hline
$t'_{2,3}$ & 0.74344 & 1.43150 & 2.03496 & 1.40707 & 0.36946 & 0.94415 & 0.93132 & 0.16985 \\
\hline
$t'_{3,3}$ & 1.35372 & 0.77597 & 1.92016 & 0.99473 & 0.81303 & 0.36259 & 0.31753 & 0.35261 \\
\hline
$t_{1,4}$ & 0.25739 & 0.22260 & 0.34850 & 0.31382 & 1.76187 & 1.65753 & 1.52861 & 1.27658 \\
\hline
$t_{2,4}$ & 1.59299 & 1.72292 & 1.31460 & 1.94008 & 0.30317 & 0.31011 & 0.21738 & 0.12688 \\
\hline
$t_{3,4}$ & 1.71721 & 1.56652 & 1.04504 & 1.46824 & 0.30507 & 0.22075 & 0.29048 & 0.42355 \\
\hline
$t'_{1,4}$ & 0.34269 & 0.25157 & 0.26769 & 0.35810 & 1.18533 & 1.19014 & 1.53603 & 1.71460 \\
\hline
$t'_{2,4}$ & 1.05712 & 1.40351 & 1.82480 & 1.16347 & 0.35652 & 0.43290 & 0.40722 & 0.11908 \\
\hline
$t'_{3,4}$ & 1.26890 & 1.30144 & 1.54555 & 1.04779 & 0.51251 & 0.25596 & 0.27451 & 0.30832 \\
\hline
$t_{1,5}$ & 0.40810 & 0.12889 & 0.30221 & 0.13525 & 1.93710 & 1.49564 & 1.37219 & 1.31394 \\
\hline
$t_{2,5}$ & 1.78749 & 1.16500 & 1.46658 & 1.36368 & 0.36629 & 0.25970 & 0.25657 & 0.47911 \\
\hline
$t_{3,5}$ & 1.77536 & 0.85654 & 1.49514 & 1.17617 & 0.73136 & 0.29417 & 0.22679 & 0.63768 \\
\hline
$t'_{1,5}$ & 0.66870 & 0.13444 & 0.31150 & 0.15842 & 1.18187 & 1.43701 & 1.46942 & 1.79038 \\
\hline
$t'_{2,5}$ & 1.13501 & 1.94727 & 1.26679 & 1.70282 & 0.37723 & 0.19528 & 0.42305 & 0.42390 \\
\hline
$t'_{3,5}$ & 1.35550 & 1.71163 & 1.36833 & 1.59387 & 0.50954 & 0.32472 & 0.23331 & 0.42474 \\
\hline
\end{tabular}
	\caption{Parameters for the corrected identity, $x$ and $y$ rotations, and $z$ rotation by
	$\frac{\pi}{2}$.}
	\label{Tab:RotParamsI}
\end{table*}

\begin{table*}
	\centering
\begin{tabular}{|c|c|c|c|c|c|c|c|c|}
\hline
Axis & $z$ & $z$ & $x+y$ & $x-y$ & $x+z$ & $x-z$ & $y+z$ & $y-z$ \\
\hline
Angle & $\pi$  & $3\pi/2$ & $\pi$ & $\pi$ & $\pi$ & $\pi$ & $\pi$ & $\pi$ \\
\hline
$j'^{\text{rot}}$ & 0.36056 & 0.17551 & 0.63069 & 0.63069 & 2.64575 & 2.22703 & 2.30059 & 2.30059 \\
\hline
$j_1^{\text{rot}}$ & 30.0000 & 30.0000 & 0.67417 & 0.67417 & 1.00000 & 30.0000 & 30.0000 & 30.0000 \\
\hline
$j_2^{\text{rot}}$ & 0.38039 & 0.26157 & 30.0000 & 30.0000 & --- & 0.87487 & 0.93444 & 0.93444 \\
\hline
$j_3^{\text{rot}}$ & 30.0000 & 30.0000 & 0.67417 & 0.67417 & --- & 30.0000 & 30.0000 & 30.0000 \\
\hline
$j_4^{\text{rot}}$ & --- & --- & --- & --- & --- & --- & --- & --- \\
\hline
$j_5^{\text{rot}}$ & --- & --- & --- & --- & --- & --- & --- & --- \\
\hline
$t_1^{\text{rot}}$ & 0.02640 & 0.03937 & 1.74236 & 0.86255 & 1.11072 & 0.05233 & 0.02672 & 0.07794 \\
\hline
$t_2^{\text{rot}}$ & 2.90256 & 3.01556 & 0.05233 & 0.05233 & --- & 1.18222 & 1.14770 & 1.14770 \\
\hline
$t_3^{\text{rot}}$ & 0.02640 & 0.03937 & 0.86255 & 1.74236 & --- & 0.05233 & 0.07794 & 0.02672 \\
\hline
$t_4^{\text{rot}}$ & --- & --- & --- & --- & --- & --- & --- & --- \\
\hline
$t_5^{\text{rot}}$ & --- & --- & --- & --- & --- & --- & --- & --- \\
\hline
$t_{1,1}$ & 0.99823 & 1.92843 & 0.08833 & 1.28673 & 0.34596 & 0.79030 & 1.60569 & 0.23302 \\
\hline
$t_{2,1}$ & 0.60853 & 0.34890 & 2.00643 & 0.68412 & 1.70052 & 1.81814 & 0.11982 & 1.50661 \\
\hline
$t_{3,1}$ & 0.23483 & 0.72084 & 1.67281 & 0.22703 & 1.56475 & 2.12906 & 0.48991 & 1.26268 \\
\hline
$t'_{1,1}$ & 1.73635 & 1.20417 & 0.12953 & 1.29749 & 0.37416 & 0.12950 & 1.42417 & 0.59732 \\
\hline
$t'_{2,1}$ & 0.26384 & 0.39221 & 1.07374 & 0.35098 & 1.41921 & 1.30677 & 0.12906 & 1.52730 \\
\hline
$t'_{3,1}$ & 0.24694 & 0.24100 & 0.89230 & 0.28196 & 1.37933 & 0.89247 & 0.26335 & 1.65259 \\
\hline
$t_{1,2}$ & 1.61084 & 1.59488 & 0.27424 & 1.42591 & 0.33335 & 0.45526 & 1.58591 & 0.18225 \\
\hline
$t_{2,2}$ & 0.25952 & 0.45350 & 1.48830 & 0.42358 & 1.34677 & 1.44068 & 0.36992 & 1.45365 \\
\hline
$t_{3,2}$ & 0.23031 & 0.13759 & 1.38719 & 0.40661 & 1.21093 & 1.24880 & 0.51445 & 1.25939 \\
\hline
$t'_{1,2}$ & 1.48086 & 1.52179 & 0.21616 & 1.39520 & 0.34354 & 0.41969 & 1.54260 & 0.18602 \\
\hline
$t'_{2,2}$ & 0.18545 & 0.34452 & 1.57898 & 0.74742 & 1.61747 & 1.45752 & 0.42459 & 1.46460 \\
\hline
$t'_{3,2}$ & 0.19828 & 0.77318 & 1.28815 & 0.47538 & 1.28189 & 1.35509 & 0.47459 & 1.25762 \\
\hline
$t_{1,3}$ & 1.31731 & 1.82596 & 0.35433 & 1.29316 & 0.28838 & 0.21660 & 1.39782 & 0.21928 \\
\hline
$t_{2,3}$ & 0.24558 & 0.86655 & 1.35594 & 0.59180 & 1.68930 & 1.46350 & 0.27381 & 1.56973 \\
\hline
$t_{3,3}$ & 0.02527 & 0.91830 & 0.77629 & 0.27573 & 1.34067 & 0.84143 & 0.92875 & 0.98139 \\
\hline
$t'_{1,3}$ & 1.81151 & 1.28752 & 0.27864 & 1.40956 & 0.25486 & 0.24862 & 1.71559 & 0.21539 \\
\hline
$t'_{2,3}$ & 0.57198 & 0.69158 & 1.78281 & 0.34085 & 1.37178 & 1.66983 & 0.19586 & 1.56537 \\
\hline
$t'_{3,3}$ & 0.98100 & 0.66898 & 1.83730 & 0.23121 & 1.02614 & 1.51736 & 0.61196 & 1.08867 \\
\hline
$t_{1,4}$ & 1.39452 & 0.83871 & 0.42834 & 1.16557 & 0.17440 & 0.33921 & 1.39500 & 0.23414 \\
\hline
$t_{2,4}$ & 0.34113 & 0.23818 & 1.37340 & 0.79534 & 1.63848 & 1.34803 & 0.24753 & 1.46441 \\
\hline
$t_{3,4}$ & 0.18321 & 0.20384 & 1.39414 & 0.36738 & 1.45126 & 1.37034 & 0.43543 & 1.39353 \\
\hline
$t'_{1,4}$ & 1.41114 & 2.28866 & 0.42670 & 1.48273 & 0.23327 & 0.28588 & 1.56864 & 0.24909 \\
\hline
$t'_{2,4}$ & 0.45458 & 0.12850 & 1.67357 & 0.44575 & 1.47534 & 1.73744 & 0.29270 & 1.45691 \\
\hline
$t'_{3,4}$ & 0.28846 & 0.33058 & 1.46928 & 0.44121 & 1.26173 & 1.56477 & 0.39172 & 1.42224 \\
\hline
$t_{1,5}$ & 1.63184 & 1.44467 & 0.14450 & 1.34021 & 0.18444 & 0.17733 & 1.29598 & 0.14930 \\
\hline
$t_{2,5}$ & 0.57060 & 0.50947 & 1.57740 & 0.39524 & 1.42391 & 1.37012 & 0.70988 & 1.64764 \\
\hline
$t_{3,5}$ & 0.73779 & 0.26562 & 1.33720 & 0.23217 & 1.18487 & 0.95442 & 0.71168 & 1.33522 \\
\hline
$t'_{1,5}$ & 1.42744 & 1.51242 & 0.13516 & 1.20153 & 0.17807 & 0.17531 & 1.81005 & 0.16545 \\
\hline
$t'_{2,5}$ & 0.49945 & 0.44739 & 1.39521 & 0.33676 & 1.69454 & 1.76074 & 0.49729 & 1.43169 \\
\hline
$t'_{3,5}$ & 0.35619 & 0.65637 & 1.10140 & 0.18018 & 1.50998 & 1.56945 & 0.53217 & 1.17211 \\
\hline
\end{tabular}
	\caption{Parameters for the dynamically-corrected $z$ rotations	by $\pi$ and $3\pi/2$, and for the rotations by
	$\hat{\vec{x}}\pm\hat{\vec{y}}$, $\hat{\vec{x}}\pm\hat{\vec{z}}$, and $\hat{\vec{y}}\pm\hat{\vec{z}}$.}
	\label{Tab:RotParamsII}
\end{table*}

\begin{table*}
	\centering
	\begin{tabular}{|c|c|c|c|c|c|c|c|c|}
	\hline
	Axis & $x+y+z$ & $x+y+z$ & $-x+y+z$ & $-x+y+z$ & $x-y+z$ & $x-y+z$ & $x+y-z$ & $x+y-z$ \\
	\hline
	Angle & $2\pi/3$ & $4\pi/3$ & $2\pi/3$ & $4\pi/3$ & $2\pi/3$ & $4\pi/3$ & $2\pi/3$ & $4\pi/3$ \\
	\hline
	$j'^{\text{rot}}$ & 1.25195 & 1.21590 & 0.26126 & 0.30530 & 1.25195 & 1.21590 & 0.30530 & 0.26126 \\
	\hline
	$j_1^{\text{rot}}$ & 1.28889 & 1.28889 & 0.34503 & 0.34503 & 1.28889 & 1.28889 & 0.34503 & 0.34503 \\
	\hline
	$j_2^{\text{rot}}$ & 30.0000 & 30.0000 & 30.0000 & 30.0000 & 30.0000 & 30.0000 & 30.0000 & 30.0000 \\
	\hline
	$j_3^{\text{rot}}$ & 1.28889 & 1.28889 & 0.34503 & 0.34503 & 1.28889 & 1.28889 & 0.34503 & 0.34503 \\
	\hline
	$j_4^{\text{rot}}$ & --- & --- & --- & --- & --- & --- & --- & --- \\
	\hline
	$j_5^{\text{rot}}$ & --- & --- & --- & --- & --- & --- & --- & --- \\
	\hline
	$t_1^{\text{rot}}$ & 1.39062 & 1.39062 & 1.16990 & 1.16990 & 0.53517 & 0.53517 & 1.79990 & 1.79990 \\
	\hline
	$t_2^{\text{rot}}$ & 0.03489 & 0.06977 & 0.06977 & 0.03489 & 0.03489 & 0.06977 & 0.03489 & 0.06977 \\
	\hline
	$t_3^{\text{rot}}$ & 0.53517 & 0.53517 & 1.79990 & 1.79990 & 1.39062 & 1.39062 & 1.16990 & 1.16990 \\
	\hline
	$t_4^{\text{rot}}$ & --- & --- & --- & --- & --- & --- & --- & --- \\
	\hline
	$t_5^{\text{rot}}$ & --- & --- & --- & --- & --- & --- & --- & --- \\
	\hline
	$t_{1,1}$ & 1.50682 & 0.23907 & 2.09919 & 2.13904 & 0.25877 & 1.32694 & 0.96335 & 0.44496 \\
	\hline
	$t_{2,1}$ & 0.43378 & 1.79424 & 0.13944 & 0.14355 & 1.37466 & 0.09653 & 0.14066 & 1.19443 \\
	\hline
	$t_{3,1}$ & 0.36195 & 1.89588 & 0.27167 & 0.24009 & 1.39970 & 0.43533 & 0.15304 & 1.30631 \\
	\hline
	$t'_{1,1}$ & 1.42114 & 0.33431 & 0.81692 & 0.88218 & 0.77381 & 1.78280 & 2.05010 & 0.26043 \\
	\hline
	$t'_{2,1}$ & 0.60324 & 1.00912 & 0.12582 & 0.11347 & 1.17007 & 0.10471 & 0.14852 & 1.26224 \\
	\hline
	$t'_{3,1}$ & 0.48662 & 1.20899 & 0.18390 & 0.22136 & 1.71261 & 0.39275 & 0.70179 & 1.27329 \\
	\hline
	$t_{1,2}$ & 1.39992 & 0.62971 & 1.49855 & 1.12516 & 0.28032 & 1.86516 & 1.48547 & 0.18537 \\
	\hline
	$t_{2,2}$ & 0.42414 & 0.77535 & 0.51557 & 0.47319 & 1.74845 & 0.39374 & 0.14925 & 1.52920 \\
	\hline
	$t_{3,2}$ & 0.27850 & 1.08828 & 0.50164 & 0.16185 & 1.71345 & 0.67478 & 0.30620 & 1.08443 \\
	\hline
	$t'_{1,2}$ & 1.21485 & 0.47335 & 1.63508 & 1.99977 & 0.67062 & 1.23605 & 1.64317 & 0.43359 \\
	\hline
	$t'_{2,2}$ & 0.42281 & 1.70063 & 0.55703 & 0.44445 & 1.12901 & 0.35980 & 0.23526 & 1.60726 \\
	\hline
	$t'_{3,2}$ & 0.22864 & 1.98788 & 0.87039 & 0.90586 & 1.40689 & 0.32649 & 0.34039 & 1.58324 \\
	\hline
	$t_{1,3}$ & 0.62378 & 0.64882 & 1.66255 & 1.55661 & 0.52880 & 1.27017 & 1.99104 & 0.18084 \\
	\hline
	$t_{2,3}$ & 1.12751 & 0.92558 & 0.24453 & 0.23475 & 1.32471 & 0.17439 & 0.22327 & 1.80390 \\
	\hline
	$t_{3,3}$ & 0.27546 & 1.36298 & 0.21519 & 0.61544 & 1.74611 & 0.18039 & 0.44019 & 1.81096 \\
	\hline
	$t'_{1,3}$ & 1.86465 & 0.39490 & 1.46305 & 1.56039 & 0.40115 & 1.78846 & 1.06507 & 0.21559 \\
	\hline
	$t'_{2,3}$ & 0.19255 & 1.60297 & 0.23591 & 0.22036 & 1.10171 & 0.15650 & 0.22715 & 1.32750 \\
	\hline
	$t'_{3,3}$ & 0.41572 & 1.75840 & 0.54332 & 0.17008 & 1.38539 & 0.46378 & 0.25086 & 0.79889 \\
	\hline
	$t_{1,4}$ & 1.12639 & 0.87640 & 1.58559 & 1.62250 & 0.41300 & 1.68174 & 1.35793 & 0.48565 \\
	\hline
	$t_{2,4}$ & 0.40157 & 1.20571 & 0.45452 & 0.39473 & 1.29101 & 0.16459 & 0.27905 & 2.07706 \\
	\hline
	$t_{3,4}$ & 0.29799 & 1.65093 & 0.92537 & 0.31961 & 1.34976 & 0.80909 & 0.05876 & 1.69366 \\
	\hline
	$t'_{1,4}$ & 1.72699 & 0.39587 & 1.51482 & 1.50224 & 0.75062 & 1.42172 & 1.63523 & 0.50811 \\
	\hline
	$t'_{2,4}$ & 0.40525 & 1.51421 & 0.54323 & 0.32635 & 1.11254 & 0.19048 & 0.20301 & 1.04977 \\
	\hline
	$t'_{3,4}$ & 0.23737 & 1.47462 & 0.36858 & 0.74476 & 1.60877 & 0.18202 & 0.61873 & 1.40129 \\
	\hline
	$t_{1,5}$ & 1.25033 & 0.28166 & 2.10165 & 2.07986 & 0.48603 & 1.22964 & 1.34488 & 0.20260 \\
	\hline
	$t_{2,5}$ & 0.36179 & 1.76795 & 0.24481 & 0.27539 & 1.22698 & 0.51213 & 0.45962 & 0.96152 \\
	\hline
	$t_{3,5}$ & 0.24346 & 1.66999 & 0.26932 & 0.44037 & 1.29670 & 0.30894 & 0.57191 & 0.49989 \\
	\hline
	$t'_{1,5}$ & 1.57572 & 0.37654 & 1.00130 & 0.99860 & 0.43093 & 1.57113 & 1.65222 & 0.23722 \\
	\hline
	$t'_{2,5}$ & 0.32856 & 1.26919 & 0.32996 & 0.31567 & 1.51202 & 0.46363 & 0.33684 & 2.15458 \\
	\hline
	$t'_{3,5}$ & 0.25086 & 1.39094 & 0.50461 & 0.40620 & 1.71000 & 0.43901 & 0.22441 & 2.01752 \\
	\hline
	\end{tabular}
	\caption{Parameters for the dynamically-corrected $\hat{\vec{x}}\pm\hat{\vec{y}}\pm\hat{\vec{z}}$ rotations.}
	\label{Tab:RotParamsIII}
\end{table*}

\end{document}